\let\orilabel\label
\documentclass[aps,amsmath,amssymb,prf,superscriptaddress]{revtex4-2} 
\let\label\orilabel

\usepackage{graphicx}
\usepackage{comment}
\usepackage{amsmath} 
\usepackage{amssymb}
\usepackage{mathtools}
\usepackage{natbib}
\usepackage{subeqnarray}
\usepackage{subfig}
\usepackage[colorlinks=true,linkcolor=blue,urlcolor=blue,citecolor=black]{hyperref}

\newcommand{\RomanNumeralCaps}[1]
\allowdisplaybreaks 

\newcommand{\RN}[1]{%
  \textup{\uppercase\expandafter{\romannumeral#1}}%
}

\newcommand\eg{\emph{e.g.}\ }
\newcommand\ie{\emph{i.e.}\ }
\newcommand\Rey{\mbox{\textit{Re}}}  

\newcommand\Wo{\mbox{\textit{Wo}}}  
\newcommand\De{\mbox{\textit{De}}}  

\newcommand{\dif}{{\mathrm{d}}}
\newcommand{\imag}{{\mathrm{i}}}

\newcommand{\norm}[1]{\left\lVert{#1}\right\rVert}

\providecommand\bnabla{\boldsymbol{\nabla}}
\providecommand\bcdot{\boldsymbol{\cdot}}
\newcommand{\lapl}[1]{\nabla^2 #1}
\newcommand{\grad}[1]{{\bnabla\! #1}}
\renewcommand{\div}[1]{{\bnabla\! \bcdot #1}}

\newcommand{\evec}{\boldsymbol{e}}
\newcommand{\fvec}{\boldsymbol{f}}

\newcommand{\nvec}{\boldsymbol{n}}

\newcommand{\qvec}{\boldsymbol{q}}

\newcommand{\rvec}{\boldsymbol{r}}
\newcommand{\svec}{\boldsymbol{s}}

\newcommand{\uvec}{\boldsymbol{u}}
\newcommand{\Uvec}{\boldsymbol{U}}

\newcommand{\Nvdis}{\textbf{N}}
\newcommand{\qvdis}{\textbf{q}}

\newcommand{\uvdis}{\textbf{u}}

\newcommand{\thetavec  }{\boldsymbol{\theta}}

\newcommand{\tauvec    }{\boldsymbol{\tau}}

\DeclareMathAlphabet\mathsfbi{OT1}{cmss}{m}{sl}

\newcommand{\Amatr}{\mathsfbi{A}}
\newcommand{\Bmatr}{\mathsfbi{B}}

\newcommand{\Ematr}{\mathsfbi{E}}
\newcommand{\Fmatr}{\mathsfbi{F}}

\newcommand{\Jmatr}{\mathsfbi{J}}

\newcommand{\Lmatr}{\mathsfbi{L}}

\newcommand{\Pmatr}{\mathsfbi{P}}

\newcommand{\Rmatr}{\mathsfbi{R}}
\newcommand{\Smatr}{\mathsfbi{S}}
\newcommand{\Tmatr}{\mathsfbi{T}}

\newcommand{\zeromatr}{\mathsfbi{0}}
\newcommand{\onematr}{\mathsfbi{1}}

\newcommand\shifthat[2]{%
  \stackengine{\Sstackgap}{$#2$}{\(\hspace{#1}\hat{}\)}{O}{l}{F}{T}{S}}
\newcommand\newhat[1]{%
\if A#1\shifthat{5.2pt}{#1}\else
\if B#1\shifthat{4.8pt}{#1}\else
\if x#1\shifthat{3.6pt}{#1}\else
\shifthat{3.4pt}{#1}
\fi
\fi
\fi
}

\newcommand\No[1][.13ex]{%
  \setbox0=\hbox{\scalebox{.7}{o}}%
  \setbox2=\hbox{n}%
  n\kern-.05em\stackengine{\dimexpr\ht0-\ht2+#1}{\belowbaseline[-\ht2]{\copy0}}%
    {\rule[-.13ex]{.7\wd0}{.13ex}}%
    {U}{c}{F}{F}{L}%
}

\usepackage{tensor} 
\graphicspath{{imgs_baseflow_pulsating/}}%

\begin{document}
\title{Characterisation of the laminar pulsatile flow in toroidal pipes}

\author{Valerio Lupi}
\email{lupi@mech.kth.se}
\affiliation{SimEx/FLOW, Engineering Mechanics, KTH Royal Institute of Technology, SE-100 44 Stockholm, Sweden}
\author{J.~Simon~Kern}
\affiliation{FLOW Turbulence Lab., Engineering Mechanics, KTH Royal Institute of Technology, SE-100 44 Stockholm, Sweden}
\author{Philipp Schlatter}
\affiliation{SimEx/FLOW, Engineering Mechanics, KTH Royal Institute of Technology, SE-100 44 Stockholm, Sweden}
\affiliation{Institute of Fluid Mechanics (LSTM), Friedrich--Alexander--Universit\"{a}t Erlangen–N\"{u}rnberg (FAU), DE-91058 Erlangen, Germany}

\begin{abstract}
This study analyses the main characteristics of the fully developed laminar pulsatile flow in a toroidal pipe as the governing parameters vary. A novel computational technique is developed to obtain time-periodic solutions of the Navier--Stokes equations. They are computed as fixed points of the system in the frequency domain via a Newton--Raphson method. Advantages and drawbacks of the adopted methodology with respect to a time-stepping technique are discussed. The unsteady component of the driving pressure gradient is found to change linearly with the pulsation amplitude, with a proportionality coefficient dependent on the pulsation frequency. Although the time-averaged streamwise wall shear stress is very close to the value in the steady case, significant fluctuations are observed within the period. Flow reversal occurs during certain time intervals in the period for high pulsation amplitudes. The analysis of the spatial structure of the unsteady component of the velocity field shows that three different flow regimes can be identified, depending on the pulsation frequency, termed quasi-steady, intermediate and plug-flow regimes.
\end{abstract}

\maketitle

\section{Introduction}
\label{sec:intro}
The flow in curved pipes is characterised by a secondary motion of Prandtl’s first kind \citep{bradshaw_1987} which increases, for instance, the heat transfer and the mixing. These effects are exploited in many industrial applications, \eg in cooling systems of nuclear reactors and food processing (see \citet{vashisth2008} for an extensive review on the topic). There are two main dimensionless parameters that govern this family of flows:~the Reynolds number $\Rey$, based on the bulk velocity $U_b$, the pipe radius $R_p$, and the fluid kinematic viscosity $\nu$, \ie $\Rey = U_b \:\! R_p/\nu$, and the curvature $\delta$, defined as the ratio between the radius of the cross-section of the pipe and the radius of curvature at the pipe centreline, \ie $\delta = R_p/R_c$. The flow in curved pipes has drawn the attention of the scientific community since the beginning of the $19^{\textrm{th}}$ century. In his pioneering works, \citet{eustice1910, eustice1911} investigated how the curvature affects the transition to turbulence and showed experimentally that secondary flow occurs in curved pipes. Later, \citet{dean1927, dean1928} derived an asymptotic solution of the incompressible Navier--Stokes equations in the limit of small curvatures and mathematically proved the existence of a secondary motion consisting of two symmetric, counter-rotating vortices, named after him. The Dean number $\De = 2 \:\! \Rey \:\! \sqrt{\delta}$ was found to be the only non-dimensional governing parameter in the solution. This result was later confirmed in the experiments of \citet{ito1959}, who measured the friction factor in curved pipes and derived an empirical relation based on the Dean number. The Dean-number similarity has been considered to hold to even higher curvatures of order 0.1 until the recent paper by \citet{canton2017}, who pointed out that it is valid only for $\delta < 10^{-6}$ and $\De < 10$.\newline
The aforementioned studies investigated laminar steady or statistically steady turbulent flows. On the other hand, several flows of biological interest are unsteady, one of the most prominent examples being the blood flow through the main arteries \citep{pedley1980}. Even industrial processes often deal with unsteady flows, as in the case of reciprocating engines \citep{kalpakli2016}. These unsteady flows can be divided into two main categories:~purely oscillatory flows (also named sinusoidal in the literature), \ie with zero mean velocity, and pulsatile flows, which have a non-vanishing mean flow rate. For the time-periodic flow in a straight pipe, a general closed-form solution can be derived in both cases \citep{womersley1955, uchida1956} since the problem becomes linear. However, the non-linearity of the Navier--Stokes equations remains in the case of curved geometries, making the analytical treatment of the problem considerably more challenging.\newline
To the best of the authors' knowledge, \citet{lyne1971} was the first to compute a solution for the oscillatory flow in a curved pipe by means of matched asymptotic expansions, valid for small curvatures and a thin Stokes layer, \ie high pulsation frequency. The analysis revealed the occurrence of a secondary flow in the Stokes layer, surrounding a steady, two-vortex motion in the core of the pipe, which has an opposite direction compared to that predicted by \citet{dean1927,dean1928} in the steady case. Hankel transforms were instead used by \citet{zalosh1973} to derive a solution in the infinitely small curvature limit but without any restriction on the pulsation frequency. \citet{smith1975} conducted an extensive investigation of the pulsatile flow in tubes of symmetric, arbitrary cross-section, covering several regimes, including steady and purely oscillatory flows.\newline
All the aforementioned analytical investigations on unsteady flows in curved pipes invoked the assumption of infinitely small curvature $\delta$. Nevertheless, flows of practical interest typically occur in geometries where this requirement is not fulfilled. In this case, a numerical or experimental approach is required due to the mathematical complexity of the problem. \citet{rabadi1980} used a first-order finite-difference scheme to perform numerical simulations of the toroidal pipe flow driven by a pulsatile pressure gradient, and they observed that the intensity of the secondary flow varies considerably within the period for low frequencies. On the other hand, they find that the time- and azimuthally-averaged axial wall shear stress is not greatly affected by either the pulsation frequency or the amplitude of the oscillating part of the pressure gradient. However, the value of the curvature considered in their work ($\delta = 0.01$) is too small to be representative of an aortic geometry, \ie $\delta \approx 0.25$ \citep{chang_tarbell_1985}. A numerical investigation of physiological pulsatile flows was carried out by \citet{chang_tarbell_1985}, who showed a qualitative agreement of the axial velocity profile with the experiments of \citet{chandran_yearwood_1981} and \citet{Yearwood1982}. The analysis of \citet{lyne1971} was later generalised by \citet{siggers_waters_2008} to take into account finite curvatures. Numerical solutions were computed using a mixed pseudo-spectral/finite-difference method to study the effect of curvature in flows driven by either purely oscillatory or pulsatile pressure gradients. The occurrence of spatially asymmetric flow fields was reported. Both periodic and quasi-periodic solutions were found. \citet{hamakiotes_berger_1990} investigated the effect of both the Reynolds number and the frequency on the flow features by numerical simulations, and observed period tripling for certain combinations of the governing parameters. It is worth mentioning that experimental studies were also conducted in parallel with the numerical simulations, analysing both purely oscillatory \citep{bertelsen1975, bertelsen_thorsen_1982, sudo1992} and physiological pulsatile flows \citep{Yearwood1982}.\newline
In the present study, the fully developed pulsatile flow through a toroidal pipe is investigated. Most of the aforementioned numerical studies did not consider low frequencies because of the substantial computational cost and storage requirement. In order to circumvent these difficulties, a novel computational technique is proposed, which is independent of the pulsation period. Solutions are computed in an extensive region of the parameter space, and the variation of the wall shear stress and the flow topology with the governing parameters is described. \newline
The remainder of the paper is organised as follows. Section~\ref{sec:problem_definition_puls} describes the problem under investigation, whereas details on the numerical method are provided in Section~\ref{sec:numerical_method_puls}. The characteristics of the laminar pulsatile flow in toroidal pipes are presented in Section~\ref{sec:results}, focusing on the wall shear stress, the friction factor and the flow topology. Eventually, Section~\ref{sec:concl} contains a summary of the paper and outlook for future work.

\section{Problem definition}\label{sec:problem_definition_puls}
\begin{figure}
  \centering
  \includegraphics{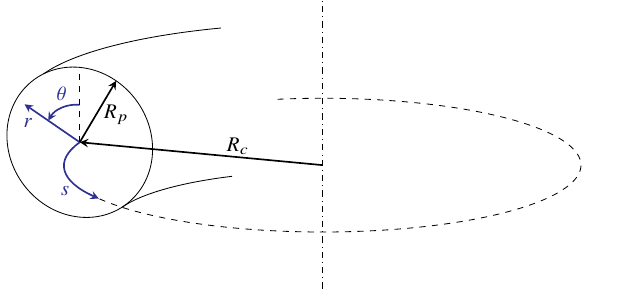}
  \caption{Sketch of a toroidal pipe with $\delta = R_p/R_c = 0.3$. The orthogonal toroidal reference system $\{s, r, \theta\}$ is also indicated.}
\label{fig:torus_ref}
\end{figure}
The laminar, pulsatile, incompressible flow of a viscous Newtonian fluid in a toroidal pipe is considered in the present study. Its dynamics is described by the incompressible Navier--Stokes equations
\begin{subeqnarray}
    & \dfrac{\partial \Uvec}{\partial t} + \big(\Uvec \bcdot \bnabla \big) \Uvec + \grad{P} - \dfrac{1}{\Rey} \lapl{\Uvec} - \fvec = 0, \\ 
    &\div{\Uvec} = 0,
\label{eq:NSE}
\end{subeqnarray}
which are made dimensionless scaling by the bulk velocity $U_b$, the radius of the cross-section of the pipe $R_p$ and the constant density of the fluid $\rho$. They are expressed in orthogonal toroidal coordinates \citep{germano1982}, sketched in Figure~\ref{fig:torus_ref}. Therefore, $\Uvec$ is the velocity vector with components $(U_s, U_r, U_{{\theta}})^T$, $P$ is the reduced pressure, and $\Rey$ is the Reynolds number based on the bulk velocity $U_b$, the radius of the pipe cross-section $R_p$ and the kinematic viscosity $\nu$. The formulation of the incompressible Navier--Stokes equations~\eqref{eq:NSE} in toroidal coordinates can be found in Appendix~\ref{sec:nonlinear_equations}. The volume force $\fvec$ in the equations~\eqref{eq:NSE} is a time-periodic forcing which drives the flow, and it mimics the effect of a streamwise pressure gradient, similarly to the steady flow investigated by \citet{canton2017} and \citet{Lupi2023a}. Note that in curved geometries, the forcing $\fvec$ is not constant across the cross-section of the pipe, as further discussed in \autoref {sec:forcing}. No-slip and impermeability boundary conditions are imposed on the pipe wall, and the flow is assumed to be homogeneous in the streamwise direction $s$.\newline
The laminar pulsatile flow in a toroidal pipe is a solution of the Navier--Stokes equations~\eqref{eq:NSE} subject to the aforementioned boundary conditions. It is both time-periodic with period $T = 2 \pi/\Omega$, with $\Omega$ being the fundamental forcing frequency, and invariant in the streamwise direction $s$. Therefore, besides the Reynolds number $\Rey$ and the curvature $\delta$, another dimensionless governing parameter is the Womersley number $\Wo = R_p \sqrt{\Omega/\nu}$, which represents the ratio of the Stokes layer thickness to the radius of the pipe \citep{lyne1971}. This solution can be numerically computed pursuing two strategies, both of which are employed in the present study. The first one is the time-stepping technique, which has been extensively used in previous studies (see \eg \citet{hamakiotes_berger_1988, hamakiotes_berger_1990} and \citet{siggers_waters_2008}) and it is described in more detail in Section~\ref{sec:nek_validation}. This method is very versatile since the same code can be used for different purposes but it may require large computational time if either the pulsation period is long ($T \propto \Rey/\Wo^2$) or a small time step is required for numerical stability. Moreover, long initial transients are often observed when adopting this method.\newline
An alternative approach is to perform a fixed-point iteration by employing the Newton--Raphson method in the frequency domain, as described in Section~\ref{sec:baseflow_solver}. Since the flow field is time-periodic, it can be expressed as a temporal Fourier series
\begin{equation}
    \Uvec(r, \theta, t) = \sum_{n \:\! = \:\! -\infty}^{\infty} \Uvec^{(n)}(r, \theta) \, e^{\imag \:\! n \:\! \Omega \:\! t}, \quad  P(r, \theta, t) = \sum_{n \:\! = \:\! -\infty}^{\infty} P^{(n)}(r, \theta) \, e^{\imag \:\! n \:\! \Omega \:\! t},
    \label{eq:uvec_fourier}
\end{equation}
and it is driven by a time-periodic volume force $\fvec$
\begin{equation}
    \fvec(r, \theta, t) = \sum_{n \:\! = \:\! -\infty}^{\infty} \fvec^{(n)}(r, \theta) \, e^{\imag \:\! n \:\! \Omega \:\! t}.
    \label{eq:fvec_fourier}
\end{equation}
Since the flow field is real-valued, one has 
\begin{equation}
    \Uvec^{(n)} = \left(\Uvec^{(-n)}\right)^*, \quad P^{(n)} = \left(P^{(-n)}\right)^*, \quad \fvec^{(n)} = \left(\fvec^{(-n)}\right)^*,
\end{equation} where the superscript $*$ denotes the complex conjugate. Substituting the expansions \eqref{eq:uvec_fourier}-\eqref{eq:fvec_fourier} in the equations~\eqref{eq:NSE} and dropping the explicit dependence on $r$ and $\theta$ for notational convenience, the Navier--Stokes equations can be written for each temporal Fourier component as
\begin{subeqnarray}
    & \imag \:\! n \:\! \Omega \, \Uvec^{(n)} + \displaystyle \sum_{n \, = \, l + m } \Big[\big(\Uvec^{(l)} \bcdot \bnabla \big) \Uvec^{(m)}\Big] + \grad{P^{(n)}} - \dfrac{1}{\Rey} \lapl{\Uvec^{(n)}} - \fvec^{(n)} = 0, \\ 
    &\div{\Uvec^{(n)}} = 0.
\label{eq:NSE_fourier}
\end{subeqnarray}
Since any time-periodic signal can be decomposed in a Fourier series, the formulation described above is valid for any waveform. In the present analysis, a purely sinusoidal forcing is considered, \ie $\fvec^{(n)} = 0, |n| \geq 2$. However, because of the intrinsic non-linearity of the Navier--Stokes equations, which introduces a coupling among the different Fourier components, a purely sinusoidal input will not result in a purely sinusoidal output in general, as opposed to linear systems. For this reason, imposing a pressure gradient with only one oscillating component gives rise to a flow field with broadband frequency content, requiring the retention of additional Fourier components.\newline
The time-periodic flow rate $Q_s$ is given by
\begin{equation}
    Q_s(t) = \int_0^{2\pi} \int_0^{R_p} U_s(r, \theta, t) \:\! r \, \dif r \, \dif \theta = \sum_{n \:\! = \:\! -\infty}^{\infty} Q^{(n)} \,  e^{\imag \:\! n \:\! \Omega \:\! t},
    \label{eq:flow_fourier}
\end{equation}
where $Q^{(n)} \ne 0$ for every $n$, in general, because of the non-linearities. The flow rate is real-valued, thus $Q^{(0)} \in \mathbb{R}$ and it can be further assumed that $Q^{(1)} \in \mathbb{R}$ without any loss of generality. Therefore, equation~\eqref{eq:flow_fourier} can be written as
\begin{equation}
    Q_s(t) = Q^{(0)} + 2 \:\! Q^{(1)} \cos({\Omega \:\! t}) + Q_{s}^{h.h.}(t) = Q^{(0)} \left( 1 + Q \cos({\Omega \:\! t})\right) + Q_{s}^{h.h.}(t),
\end{equation}
where $Q_{s}^{h.h.}(t)$ accounts for the higher harmonic components and $Q = 2 \, Q^{(1)}/Q^{(0)}$ is the relative amplitude of the pulsation and represents an additional non-dimensional governing parameter of the flow. For fixed $Q^{(0)}$ and $Q$, the total flow rate varies for different governing parameters since the amplitude of $Q_{s}^{h.h.}(t)$ depends on them. However, for most cases investigated in this study, the flow rate of the higher harmonics does not contribute significantly to the total one.\newline
For the reasons stated above, the oscillating components of the forcing $\fvec$, the flow rate $Q_s$ and the velocity $\Uvec$ cannot all have a purely sinusoidal time-dependence simultaneously. Imposing a purely sinusoidal variation for one of the quantities results in a broadband frequency content for the other two. Therefore, the prescription of the driving mechanism of the flow is of paramount importance for the correct definition of the problem.

\subsection{Forcing mechanism} \label{sec:forcing}
The driving force is not uniform over the cross-section. In order to mimic the effect of a streamwise pressure gradient, its spatial structure needs to be chosen such that the difference between two points at the same location in the cross-section but at different streamwise stations depends only on their streamwise distance $\Delta s$. Therefore, following the same reasoning as \citet{canton2017}, the forcing in toroidal coordinates can be formulated for each temporal Fourier component as
\begin{equation}
    \fvec^{(n)} = \dfrac{F^{(n)}}{1 + \delta \, r \sin({\theta})} \, \evec_s,
    \label{eq:forcing_shape}
\end{equation}
where $\evec_s$ is the unit vector in the streamwise direction and the forcing amplitude $F^{(n)}$ is computed to have a desired flow rate $Q^{(n)}_{\textrm{target}}$ by solving for 
\begin{equation}
Q^{(n)}\bigl( \fvec^{(n)} \bigr) - Q_{\textrm{target}}^{(n)} = 0.
\label{eq:forcing_bulk}
\end{equation}
For the problem to be fully defined, the target flow rate needs to be set for all Fourier components such that $\fvec^{(n)} \ne 0$. The non-linear relation between $\fvec^{(n)}$ and $Q^{(n)}$ is governed implicitly by the equations~\eqref{eq:NSE_fourier} and thus needs to be determined iteratively.

\section{Numerical methods}\label{sec:numerical_method_puls}
In this section, the numerical approaches for discretising and eventually solving the governing equations are discussed. As stated earlier, two different approaches are introduced: the fixed-point iteration and a time-stepping method. Solutions obtained with the two strategies are eventually compared in Section~\ref{sec:validation}.

\subsection{Fixed-point iteration}\label{sec:baseflow_solver}
The original problem is collapsed onto a Poincar\'{e} section by writing the Navier--Stokes equations in the frequency domain. Therefore, the laminar time-periodic solution corresponds to a fixed point of the Poincar\'{e} map. It is computed numerically by solving the system~\eqref{eq:NSE_fourier} together with equation~\eqref{eq:forcing_bulk} using the Newton--Raphson method. Although a similar framework as in \citet{Lupi2023a} is adopted, modifications are required to deal with the solution of the equations in the frequency domain. Indeed, the interactions between the Fourier components need to be taken into account. Writing $\mathcal{L}_1$ as the continuous operator for the linear terms in the equations~\eqref{eq:NSE_fourier}, $\mathcal{L}_2(\uvec)$ as the operator linearising the convective terms about a given state $\qvec = (\uvec, p)^T$, and introducing the restriction operator $\mathcal{R}$ (identity for the velocity, zero for the pressure), the Newton--Raphson method at the iteration $k$ can be written as
\begin{equation}
    \Big(\imag \:\! n \:\! \Omega \, \mathcal{R} + \mathcal{L}_1 \Big) \, \Delta \qvec^{(n)}_{k} + \sum_{n \, = \, l + m} \Big[\mathcal{L}_2\Big(\uvec^{(l)}_{k-1}\Big)  \, \Delta \qvec^{(m)}_{k}\Big] =  -\mathcal{N}\Big(\qvec_{k-1}^{(n)}\Big),
\label{eq:linNSE_fourier}
\end{equation}
where $\mathcal{N}(\qvec^{(n)})$ is a shorthand for the non-linear Navier--Stokes equations~\eqref{eq:NSE_fourier}, $\qvec^{(n)} = (\uvec^{(n)}, p^{(n)})^T$ and $\Delta \qvec^{(n)}_{k} = \qvec^{(n)}_{k} - \qvec^{(n)}_{k-1}$. The expressions for $\mathcal{L}_1$ and $\mathcal{L}_2(\uvec)$ in toroidal coordinates are reported in Appendix~\ref{app:linear_operators}.

\subsubsection{Spatial and temporal discretisation}
The numerical solution of the equations~\eqref{eq:forcing_bulk}-\eqref{eq:linNSE_fourier} is carried out on a two-dimensional domain represented by the pipe cross-section since the flow is homogeneous in the streamwise direction $s$. The in-house developed \textsc{Matlab}$^{\circledR}$ code used by \citet{Lupi2023a} is adapted for taking into account pulsating flows. The spatial discretisation is unchanged, employing Fourier basis functions in the azimuthal direction ${\theta}$ and Chebyshev polynomials in the radial direction $r$, and relying on the \textsc{Matlab}$^{\circledR}$ \texttt{DMSUITE} by \citet{dmsuite}. The equations are discretised at $N_1 = n_r \times n_{\theta}$ collocation points, where $n_r$ and $n_{{\theta}}$ represent the number of grid points in the radial and azimuthal directions, respectively. Since the computational grid is collocated, spurious pressure modes arise \citep{canuto1988}. Their treatment is carried out as described in more detail by \citet{gustavsson2003}.\newline
The continuous operators $\mathcal{L}_1, \mathcal{L}_2(\uvec)$ and $\mathcal{R}$ are replaced with their discrete counterparts, denoted $\Lmatr_1$, $\Lmatr_2(\uvdis)$ and $\Rmatr$, respectively, and having the following macrostructure
\begin{subequations}
\label{eq:discr_oper}
\begin{align}
     \Lmatr_1 &= \begin{bmatrix}
             \Lmatr_{1, s s}       & \Lmatr_{1, s r}       & \Lmatr_{1, s \theta}             & \Lmatr_{1, s p} \\
             \Lmatr_{1, r s}       & \Lmatr_{1, r r}       & \Lmatr_{1, r \theta}             & \Lmatr_{1, r p} \\
             \Lmatr_{1, \theta s} & \Lmatr_{1, \theta r} & \Lmatr_{1, \theta \theta}       & \Lmatr_{1, \theta p} \\
             \Lmatr_{1, p s}       & \Lmatr_{1, p r}       & \Lmatr_{1, p \theta}             & \zeromatr_{N_1} \\
         \end{bmatrix} \in \mathbb{C}^{2N \times 2N} \:, \label{eqn:L1} \\[4pt]
     \Lmatr_2(\uvdis) &= \begin{bmatrix}
             \Lmatr_{2, s s}(\uvdis)          & \Lmatr_{2, s r}(\uvdis)        & \Lmatr_{2, s \theta}(\uvdis)  & \zeromatr_{N_1} \\
             \Lmatr_{2, r s}(\uvdis)          & \Lmatr_{2, r r}(\uvdis)        & \Lmatr_{2, r \theta}(\uvdis)  & \zeromatr_{N_1} \\
             \Lmatr_{2, \theta s}(\uvdis)    & \Lmatr_{2, \theta r}(\uvdis)  & \Lmatr_{2, \theta \theta}(\uvdis)  & \zeromatr_{N_1} \\
             \zeromatr_{N_1}   & \zeromatr_{N_1} & \zeromatr_{N_1} & 
\zeromatr_{N_1} \\
         \end{bmatrix} \in \mathbb{C}^{2N \times 2N},  \label{eqn:L2} \\[4pt]
     \Rmatr &= \mbox{diag}(\onematr_{N_1}, \onematr_{N_1}, \onematr_{N_1}, \zeromatr_{N_1}),
\end{align}
\end{subequations}
where $\onematr_n$ and $\zeromatr_n$ are the identity and zero matrix of order $n$, respectively, and $2 N = 4 N_1$. In each operator, the first three rows discretise the components of the momentum equation and the last row corresponds to the discretised incompressibility constraint. The zero sub-blocks are due to the fact that the pressure is not coupled to itself and is not involved in the convective terms.\newline
In the numerical solution of the system~\eqref{eq:linNSE_fourier}, the temporal Fourier series is truncated at some $K_b \in \mathbb{N}$, such that $N_b = 2 \, K_b + 1$ frequency components are retained. Introducing the discrete operators defined in the equations~\eqref{eq:discr_oper} and gathering all the temporal Fourier components in one vector, we obtain
\begin{equation}
    \Delta \qvdis =  \left[\Delta \qvdis^{(-K_b)}, \ldots, \Delta \qvdis^{(K_b)}\right]^T, \quad \qvdis =  \left[\qvdis^{(-K_b)}, \ldots, \qvdis^{(K_b)}\right]^T,
\end{equation}
and the Newton--Raphson method reads
\begin{equation}
    \Fmatr_{N_b} \, \Delta \qvdis = -\Nvdis(\qvdis),
    \label{eq:newton_method}
\end{equation}
where $\Nvdis(\qvdis)$ represents the discretised non-linear Navier--Stokes equations in the frequency domain~\eqref{eq:NSE_fourier}. The discrete Jacobian $\Fmatr_{N_b}$ can be expressed using the Kronecker product $\otimes$ as
\begin{equation}
    \Fmatr_{N_b} = \Rmatr_f \cdot \left[ \Pmatr \otimes \onematr_{2N} \right] + \left[ \onematr_{N_b} \otimes \Lmatr_1 \right] + \sum_{n \, = \, -K_b}^{K_b} \left[ \onematr^{(n)}_{N_b} \otimes \Lmatr_2\big(\textbf{U}^{(n)}\big) \right]
\end{equation}
where $\Pmatr = \imag \:\! \Omega \cdot \mbox{diag}(-K_b, \ldots, K_b)$  and $\onematr^{(n)}_k$ is the shift matrix of order $k$ with ones on the $n^{\textrm{th}}$ sub/super-diagonal.\newline
\citet{siggers_waters_2008} reported the existence of stable asymmetric solutions for the oscillatory flow in a toroidal pipe. For $\delta = 0$, \ie a straight pipe, these solutions are found by continuation also in the pulsatile regime, up to a certain Reynolds number. However, it is not clear if they also exist for pulsatile flows through toroidal pipes and whether they are stable or not. In the present study, following \citet{hamakiotes_berger_1988, hamakiotes_berger_1990}, the laminar pulsatile flow is assumed to be symmetric along the equatorial plane of the torus. This allows the restriction of the computational domain to only half the cross-section of the pipe, hence reducing the number of points in the azimuthal direction to $n_{\theta s} = n_\theta/2$. The symmetrisation is introduced at the level of the operators $\Lmatr_1$, $\Lmatr_2(\uvdis)$ and $\Rmatr$, before the discrete Jacobian is assembled. For each operator $\Lmatr_i$, its symmetrised version $\Lmatr_{i,s}$ is computed as
\begin{equation}
  \Lmatr_{i,s} = \Tmatr_s \:\! \Lmatr_i \:\! \Tmatr\in \mathbb{C}^{N \times N}, \quad i = 1,2,
\end{equation}
where $\Tmatr_s \in \mathbb{R}^{N \times 2N}$ and $\Tmatr \in \mathbb{R}^{2N \times N}$ are the symmetrisation operators defined in \ref{sec:TTs}. The symmetrisation of the flow field vector is required as well, \ie
\begin{equation}
   \qvdis_s = \Tmatr_s \, \qvdis \in \mathbb{C}^{N}.
    \label{eq:vs_disc}
\end{equation}
At each iteration of the Newton--Raphson method, the linear system in equation~\eqref{eq:newton_method} is solved using a direct method. An improvement of the computational performance might be obtained using an iterative method, \eg GMRES, preferably together with preconditioning techniques. However, if the size of the problem is very large, the explicit construction of the discrete Jacobian is unfeasible because of the memory requirements, and a Jacobian-free Newton--Krylov method is preferable.\newline
The solution is considered converged when 
\begin{equation}
    \sum_{n \, = \, -K_b}^{K_b} \frac{1}{3 \, N_1} \, \norm{\Rmatr \, \Nvdis\big(\qvdis^{(n)}\big)}_{L_2} < 10^{-9} \quad \mbox{and} \quad \left| Q_s^{(n)} - Q_{s,\textrm{target}}^{(n)} \right|  < 10^{-9}, \ n \leq 1.
\end{equation}
Alternatively, the algorithm is stopped if it diverges or the maximum number of iterations $n_{\textrm{max}} = 50$ is reached. The Newton--Raphson method is only locally convergent. In the present work, the steady state solution ($K_b = 0$) for a given curvature $\delta$ is computed starting from the solution of the Stokes equations, and the Reynolds number is progressively increased. For each value of the Womersley number, the laminar steady solution is provided as an initial guess for computing the solution with $K_b = 1$ at the same $\delta$ and $Re$. The pulsation amplitude $Q$ is then gradually increased until either the maximum value of $Q = 0.5$ is reached or a converged solution cannot be obtained. This approach is repeated for different values of $K_b$, initialising the algorithm with the solution retaining a lower number of Fourier components.\newline
As stated previously, the major drawback of the present method is that the number of Fourier components required to capture all relevant non-linear interactions is not known a priori. Therefore, the influence of the truncation of the Fourier series needs to be analysed a posteriori, as explained further in Section~\ref{sec:reso_Kb}. Moreover, the convergence of the fixed-point iteration depends on both the quality of the initial guess and the stiffness of the problem. These limitations are likely a reason why some solutions in the parameter space cannot be computed using the current approach.

\subsection{Time-stepping method} \label{sec:nek_validation}
The time-stepping approach is carried out with the spectral-element code \texttt{nek5000} \citep{nek5000}. The computational domain consists of a portion of a three-dimensional toroidal pipe discretised using hexahedral elements. Within each element, Lagrangian interpolants of order $N_p$ built on Gauss--Lobatto--Legendre (GLL) nodes are used to represent the velocity, whereas the pressure is expressed as a linear combination of Lagrangian basis functions of order $N_p - 2$ based on Gauss--Legendre (GL) grid points. For the present study, polynomial order $N_p = 7$ is chosen. The Navier--Stokes equations are formulated in Cartesian coordinates and they are integrated in time using a third-order backward differentiation formula (BDF3), but an extrapolation scheme of order three (EXT3) is used for computing the non-linear term at the next time step. Periodic boundary conditions are applied in the streamwise direction, taking care of the rotation of the velocity components. The volume force driving the flow can be directly imposed or, alternatively, it can be derived by prescribing a desired flow rate. The latter case represents the time-varying counterpart of the approach employed by \citet{noorani2015}. The convergence of the time-periodic solution is tested by requiring that
\begin{equation}
    \norm{\Uvec(t_0 + T) - \Uvec(t_0)}_E < 10^{-10},
\end{equation}
where $\norm{\bcdot}_E$ represents the standard energy norm.\newline
The main advantage of the time-stepping approach consists in taking into account all the temporal frequency components of the flow, without the need for truncation. However, periodicity is not a priori guaranteed, and unstable solutions cannot be computed unless a way to stabilise them is implemented. Instead, in the fixed-point iteration method in the frequency domain, the solution is periodic in time by construction, and therefore unstable periodic orbits can be computed.  

\subsection{Temporal resolution study} \label{sec:reso_Kb}
\begin{figure}
   \centering
   \subfloat{\includegraphics[]{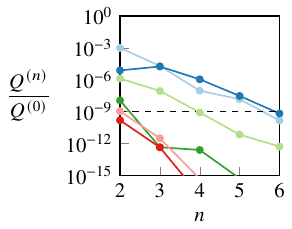}} \label{fig:d0p01_mferr_Q0p1} 
   \subfloat{\includegraphics[]{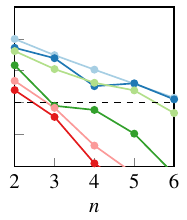}} \label{fig:d0p1_mferr_Q0p1}
   \subfloat{\includegraphics[]{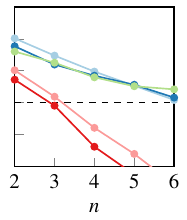}} \label{fig:d0p3_mferr_Q0p1}\\ \vskip 3pt
   \hspace{1.7cm}
   \subfloat{\includegraphics[]{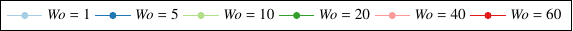}}
   \caption{Mass flow rate of the Fourier components of the solution computed with $K_b = 6$ at $\Rey = 500$, $Q = 0.1$ and different Womersley numbers. (\textit{Left}) $\delta = 0.01$, (\textit{middle}) $\delta = 0.1$, (\textit{right}) $\delta = 0.3$.}
   \label{fig:res_mferr_Q0p1}
\end{figure}

\begin{figure}
   \centering
   \includegraphics{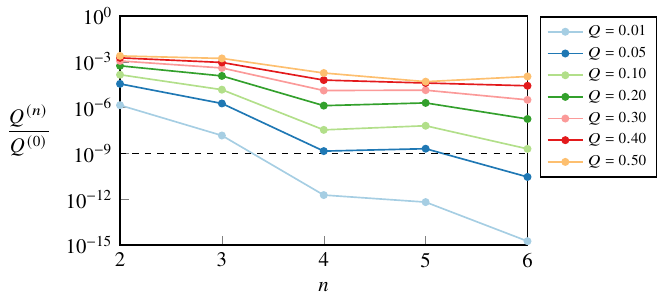}
   \caption{Mass flow rate of the Fourier components of the solution computed with $K_b = 6$ at $\Rey = 500$, $\delta = 0.1$, $\Wo = 5$ and different values of $Q$.}
   \label{fig:d0p1_mferr_Wo_5} 
\end{figure}

Referring to the procedure described in Section~\ref{sec:baseflow_solver}, no further extension of the temporal Fourier series is performed if the solution for a given $K_b$ satisfies also the problem for a higher value of $K_b$. However, for some of the considered values of $Q$, this approach would require to compute the laminar pulsatile solution for large values of $K_b$, which is not affordable. Therefore, in order to assess the necessary number of Fourier components to be retained in the solution of the system, the variation of relevant flow quantities with $K_b$ is investigated for different values of $\Wo$ and $Q$. The mass flow rate of the Fourier components of the solution computed with $K_b = 6$ for $\Rey = 500$, $\delta = [0.01, 0.1, 0.3]$, $Q = 0.01$, $\Wo = [1, 5, 10, 20, 40, 60]$ is shown in Figure~\ref{fig:res_mferr_Q0p1}. The number of Fourier components that need to be retained is significantly affected by the Womersley number. Indeed, the problem requires higher resolution in the frequency domain for $1 \lessapprox \Wo \lessapprox 20$, indicating that non-linear interactions among the Fourier components are strong in this range of Womersley numbers. On the other hand, they play a smaller role for $\Wo \gtrapprox 40$. This result is conjectured to be related to the different flow patterns occurring for different Womersley numbers, as later described in Section~\ref{sec:flow_regimes}. The strength of the non-linear interactions among the Fourier components of the solution increases with the pulsation amplitude, resulting in a broader frequency spectrum, as displayed in Figure~\ref{fig:d0p1_mferr_Wo_5}. This is likely the reason why the method does not reach convergence for $\delta = 0.3$, $\Wo = 20$ and $Q \geq 0.1$.\newline
It is worth noting that, while computing a solution with $K_b = 6$ is possible for the present study, it is not feasible for cases requiring considerably higher spatial resolution, \eg at larger Reynolds numbers. However, the results herein presented show that, for a given Womersley number and pulsation amplitude, the mass flow rate decreases almost exponentially with the flow harmonics. An exponential extrapolation can then be performed to assess the required resolution in the frequency domain. However, at intermediate Womersley numbers, $Q^{(n)}$ does not decay exponentially with $n$ and it exhibits a non-monotonic trend. This is ascribed to the fact that the temporal Fourier components of the velocity have considerably different spatial structures for these values of $\Wo$. It should be noted that the flow rate takes into account only the streamwise velocity component $U_s$. Therefore, a small value of $Q^{(n)}$ does not guarantee that the in-plane velocity components are negligible. An alternative criterion for considering the effect of $U_r$ and $U_{\theta}$ is to analyse the magnitude of the non-linear term $(\Uvec^{(l)} \bcdot \bnabla) \Uvec^{(m)}$ for $ l + m > K_b$. 
\subsection{Validation} \label{sec:validation}
\begin{figure}
   \centering
   \includegraphics{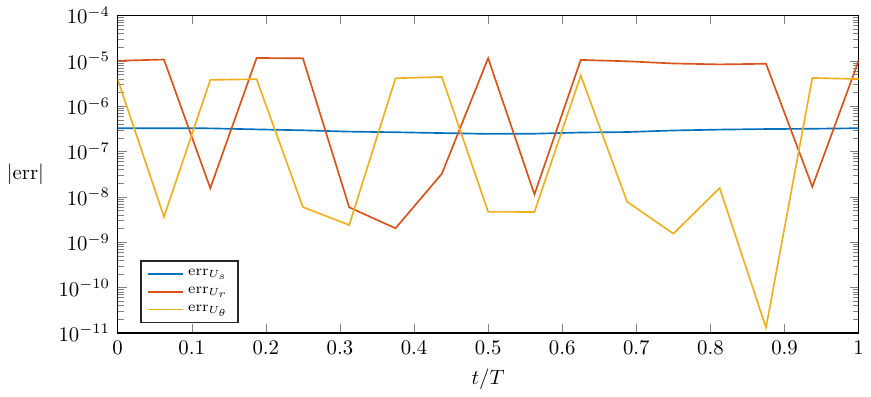}
   \caption{Median value of the relative error on the cross-section between the velocity components computed with the fixed-point iteration method and the time-stepping algorithm. The pulsatile flow at $\delta = 0.3$, $\Rey = 500$, $\Wo = 20$ and $Q = 0.05$ is considered.}
   \label{fig:err_nek_matlab} 
\end{figure}

A validation of the fixed-point iteration method is performed by comparing the laminar time-periodic solution computed for $K_b = 6$ with the one obtained employing the time-stepping code, in both cases for $\delta = 0.3$, $\Rey = 500$, $\Wo = 20$ and $Q = 0.01$. For the purpose of validation, the same forcing as the one derived from the fixed-point iteration algorithm is imposed in the time-stepping code. The median value of the relative error on the cross-section is shown in Figure~\ref{fig:err_nek_matlab} for the three velocity components over the period of pulsation. A satisfactory agreement is observed between the two methodologies. The median is used as a measure of agreement since it is less sensitive to outliers occurring where the relative error is high because of low absolute values of the velocity and large interpolation errors.

\section{Pulsatile flow characteristics} \label{sec:results}
The main features of the laminar pulsatile flow in a toroidal pipe are presented and discussed in this section. Although the numerical methods employed in the current work can compute a solution for a large range of the governing parameters, the following analysis focuses on laminar pulsatile flows at $\Rey = 500$, $\delta = [0.01, 0.1, 0.3$], and various values of $\Wo$ and $Q$. These values of $\delta$ are chosen since they are illustrative of a relevant range of curvatures. The choice of presenting only one value of the Reynolds number is instead justified by the fact that this parameter does not have a significant influence on the overall structure of the flow \citep{Kern2023}. All the results presented in the following refer to solutions computed setting $K_b = 6$ unless differently stated. Furthermore, for higher values of $\Rey$, the laminar pulsatile solution may become unstable \citep{Kern2023}, thus losing its relevance.

\subsection{Volume forcing} \label{sec:forcing_scaling}
\begin{figure}
\centering
\includegraphics{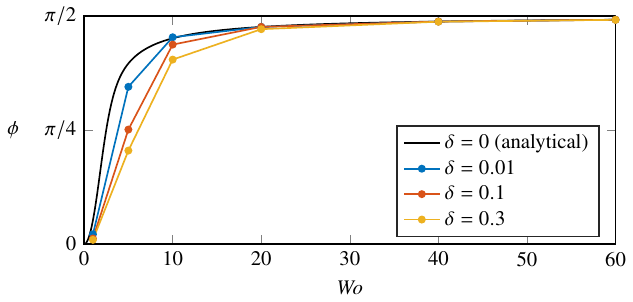}
   \caption{Phase lag between the volume force and the flow rate as a function of the Womersley number for $\Rey = 500$, $Q = 0.05$ and different curvatures.}
   \label{fig:phase_lag} 
\end{figure}

\begin{figure}
   \centering
   \subfloat{\includegraphics[]{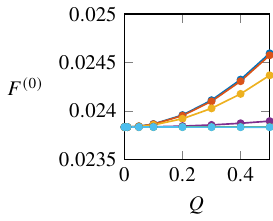}} \label{fig:d0p01_f0} 
   \subfloat{\includegraphics[]{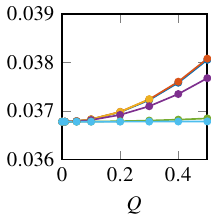}} \label{fig:d0p1_f0} 
   \subfloat{\includegraphics[]{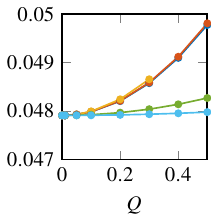}} \label{fig:d0p3_f0}\\ \vskip 3pt
   \hspace{1.9cm}
   \subfloat{\includegraphics[]{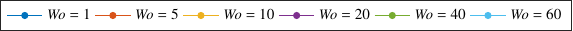}} 
   \caption{Steady components of the forcing for $\Rey = 500$ and different values of $\Wo$ and $Q$. (\textit{Left}) $\delta = 0.01$, (\textit{middle}) $\delta = 0.1$, (\textit{right}) $\delta = 0.3$.}
   \label{fig:fs0_Q}
\end{figure}

\begin{figure}
   \centering
   \subfloat{\includegraphics[]{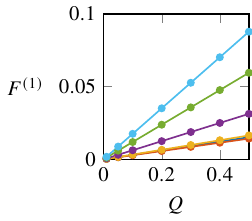}} \label{fig:d0p01_f1} 
   \subfloat{\includegraphics[]{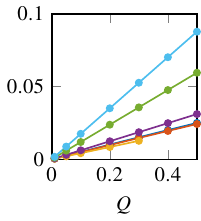}} \label{fig:d0p1_f1}
   \subfloat{\includegraphics[]{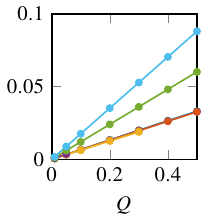}} \label{fig:d0p3_f1}\\ \vskip 3pt
   \hspace{1.7cm}
   \subfloat{\includegraphics[]{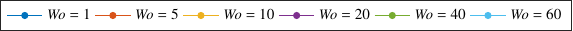}} 
   \caption{First Fourier components of the forcing for $\Rey = 500$ and different values of $\Wo$ and $Q$. (\textit{Left}) $\delta = 0.01$, (\textit{middle}) $\delta = 0.1$, (\textit{right}) $\delta = 0.3$.}
   \label{fig:fs1_Q}
\end{figure}

It is well-known that, for the laminar pulsatile flow in a straight pipe, a phase lag exists between the streamwise pressure gradient and the resulting flow field, and that it approaches $\pi/2$ as $\Wo \rightarrow \infty$ \citep{womersley1955, uchida1956}. This result can be derived analytically as follows. The momentum equation in the streamwise direction reads
\begin{equation}
    \dfrac{\partial U^{(n)}}{\partial t} = \dfrac{1}{\Rey} \left(\dfrac{1}{r} \dfrac{\partial U^{(n)}}{\partial r} + \dfrac{\partial^2 U^{(n)}}{\partial r^2}\right) + f^{(n)}, 
    \label{eq:NS_straight}
\end{equation}
where $f^{(n)} = - \partial P/ \partial z$. Following \citet{womersley1955,uchida1956}, the expression for the streamwise velocity profile can be written as
\begin{equation}
    U^{(n)}(r,w) = \begin{cases}
                    2 \:\! (1-r^2) ,  &\quad \text{if  } n =  0, \\[10pt]
                    \dfrac{1}{B}\left( \dfrac{J_0( \sqrt{-\imag} \:\! r \:\! w)}{A} - 1 \right) & \quad \text{otherwise},
                \end{cases}
\end{equation}
with $w = \sqrt{n} \, \Wo$ and the constants $A$ and $B$ given by
\begin{equation}
    A = J_0(\sqrt{-\imag} \:\! w), \quad
    B = \dfrac{2}{\sqrt{-\imag} \:\! w} \dfrac{J_1(\sqrt{-\imag} \:\! w)}{J_0(\sqrt{-\imag} \:\! w)} - 1,
\end{equation}
where $J_0$ and $J_1$ are Bessel functions of the first kind. Substituting these expressions into equation~\eqref{eq:NS_straight}, noting the derivative rules for Bessel functions
\begin{equation}
    \dfrac{\partial J_0(r)}{\partial r}  = - J_1(r), \quad 
    \dfrac{\partial^2 J_0(r)}{\partial r^2} = - \dfrac{\partial J_1(r)}{\partial r} = - J_0(r) + \dfrac{J_1(r)}{r},
\end{equation}
an expression for the steady and unsteady components of the forcing can be obtained as
\begin{equation}
    \dfrac{\pi R_p^2}{Q^{(n)}} \:\! f^{(n)} = \begin{cases}
                    \dfrac{8}{\Rey},  &\quad \text{if  } n =  0, \\[10pt]
                    - \dfrac{\imag \:\! n \:\! \Omega}{B} & \quad \text{otherwise}.
                \end{cases}
\end{equation}
Given that 
\begin{equation}
    \frac{Q^{(0)}}{\pi R^2} \equiv 1, \quad Q = \frac{2 \:\! Q^{(1)}}{Q^{(0)}},
\end{equation}
the unsteady forcing components can be written as
\begin{equation}
    f^{(n)} = - \dfrac{Q}{2} \, \dfrac{\imag \:\! n \:\!\Omega}{B}, \quad n \ne 0.
\end{equation}
Figure~\ref{fig:phase_lag} shows the phase lag $\phi$ computed in the current work for $\delta = [0.01, 0.1, 0.3]$, $\Rey = 500$, $\Wo = [1,5,10,20,40,60]$ and $Q = 0.05$ together with the analytical solution for the straight pipe. For a given Womersley number, it is observed that the relative variation of the phase lag with the pulsation amplitude is at most of $2\%$. Therefore, only data for one value of $Q$ are shown.
\newline
In the toroidal pipe, as the pulsation frequency increases, the phase lag approaches the asymptotic value of $\pi/2$, with a rate that increases with decreasing curvature. In the range $5 \lessapprox \Wo \lessapprox 10$, the curvature has a strong effect on the phase lag, whereas its influence is significantly reduced for higher frequencies. These findings are not in agreement with those by \citet{lin_tarbell_1980}, who observed that the phase lag between the flow field and the pressure gradient converges to the asymptotic limit of 80.4\textdegree\ as $\Wo \rightarrow \infty$. However, their conclusion is based on data for $\delta = 0.05$ and $\Wo < 46$. On the other hand, \citet{hamakiotes_berger_1988}, studying the pulsatile flow for $\delta = 1/7$ and $\Wo = 15$, found a phase lag of approximately $90^{\circ}$, regardless of the Reynolds number. Nevertheless, the authors did not report the behaviour of the phase lag with either the frequency or the curvature. \newline
The current results suggest that the non-linear interactions play a role also in determining the phase lag between the streamwise pressure gradient and the velocity. As the pulsation amplitude increases, the steady component of the forcing becomes larger than the value in the steady case ($Q = 0$). The increase goes almost quadratically with $Q$, as it can be observed in Figure~\ref{fig:fs0_Q}. Note that this quadratic dependence is more pronounced for lower values of $\Wo$, whereas it is almost negligible at high frequencies. This behaviour is expected since the non-linearity is quadratic. For high values of $\Wo$, the steady component is not significantly affected by the unsteady one, which eventually dominates the dynamics, since non-linear interactions are weaker. For a fixed value of the Womersley number, the amplitude of the unsteady component of the forcing $F^{(1)}$ increases linearly with $Q$, as shown in Figure~\ref{fig:fs1_Q} for different values of the curvature. The proportionality coefficient is found to increase monotonically with $\Wo$, confirming that the unsteady component becomes more dominant as the frequency of the pulsation increases. The observed scalings, although not analytical, can be used to extrapolate the forcing to be provided as input in the time-stepping method for different values of the pulsation amplitude. 


\subsection{Friction factor and wall shear stress}
\begin{figure}
    \centering
    \hspace{-0.325cm}
   \subfloat{\includegraphics[]{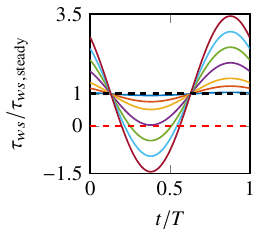}} \label{fig:d0p01_tauws_Wo40} 
   \subfloat{\includegraphics[]{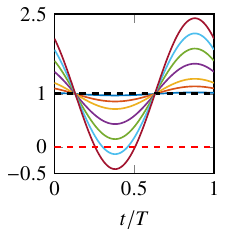}} \label{fig:d0p1_tauws_Wo40}
   \subfloat{\includegraphics[]{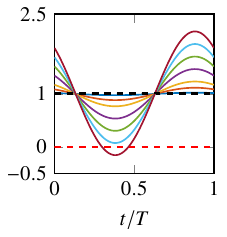}} \label{fig:d0p3_tauws_Wo40}\\ \vskip 3pt
   \hspace{1.15cm}
   \subfloat{\includegraphics[]{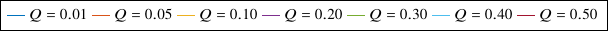}} 
   \caption{Streamwise component of the wall shear stress for $\Rey = 500$, $\Wo = 40 $ and different values of $Q$ at the outer wall ($\theta = \pi/2$). (\textit{Left}) $\delta = 0.01$, (\textit{middle}) $\delta = 0.1$, (\textit{right}) $\delta = 0.3$.}
   \label{fig:tauws_Wo40}
\end{figure}

\begin{figure}
    \centering
   \includegraphics{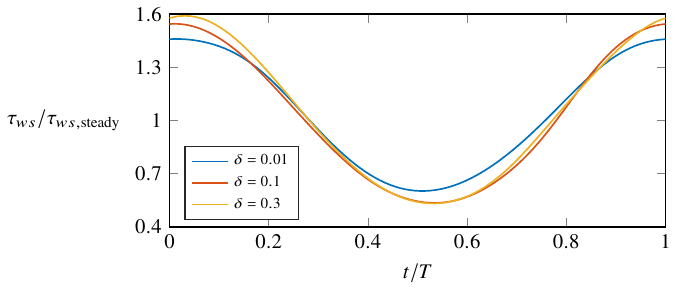}
   \caption{Streamwise component of the shear stress for $\Rey = 500$, $\Wo = 5 $ and $Q = 0.4$. $\delta = [0.01, 0.1, 0.3]$ at the outer wall ($\theta = \pi/2$).}
   \label{fig:tauws_Wo5_Q0p4} 
\end{figure}

\begin{figure}
    \centering  
    \includegraphics{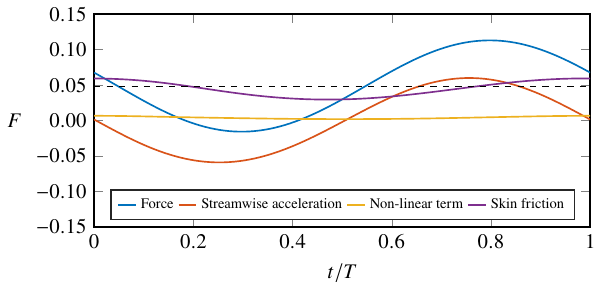}
   \caption{Balance of forces for $\Rey = 500$, $\Wo = 10 $, $Q = 0.3$, $\delta = 0.3$. The black dashed line represents the forcing amplitude in the steady case for the same values of the Reynolds number and curvature.}
   \label{fig:force_bal_d0p3_Wo10_Q0p3} 
\end{figure}

The wall shear stress is an important quantity of interest for both industrial and biological applications of pulsatile flows in curved pipes. It is relevant from a structural point of view and it affects the power needed to drive the flow in a pipeline. It is also thought that the wall shear stress distribution plays a role in the formation of atherosclerotic plaques \citep{berger1983}.\newline
For incompressible flows, the deviatoric stress tensor is defined as
\begin{equation}
    \Smatr = 2 \mu \Ematr,
\end{equation}
where $\Ematr$ represents the strain-rate tensor. The shear stress vector at the wall can be computed as
\begin{equation}
    \tauvec_w = \left. \Smatr \right \rvert_{ \:\! r \:\! = R_P} \cdot \nvec_w,
\end{equation}
where $\nvec_w$ is the wall-normal unit vector. Since the flow is homogeneous in the streamwise direction and the velocity components are zero at the wall, many terms vanish when evaluating the shear stress vector at the wall. Its streamwise and azimuthal components reduce to
\begin{equation}
   \tau_{w s} = \mu \left. \dfrac{\partial u_s}{\partial r}\right \rvert_{\, r \, = \, R_p} , \quad \tau_{w \theta} = \mu \left. \dfrac{\partial u_{\theta}}{\partial r} \right \rvert_{\, r \, = \, R_p}.
\end{equation}
\begin{figure}
   \centering
   \subfloat{\includegraphics[]{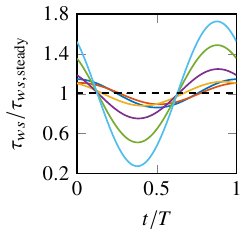}} \label{fig:d0p01_tauws} 
   \subfloat{\includegraphics[]{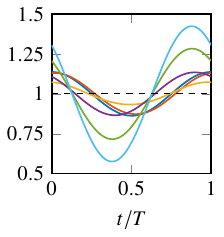}} \label{fig:d0p1_tauws}
   \subfloat{\includegraphics[]{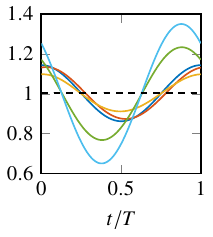}} \label{fig:d0p3_tauws}\\ \vskip 3pt
   \hspace{1.15cm}
   \subfloat{\includegraphics[]{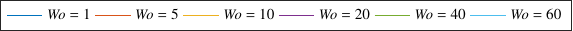}} 
   \caption{Streamwise component of the wall shear stress for $\Rey = 500$, $Q = 0.1$ and $\Wo = [1, 5, 10, 20, 40, 60]$ at the outer wall ($\theta = \pi/2)$. (\textit{Left}) $\delta = 0.01$, (\textit{middle}) $\delta = 0.1$, (\textit{right}) $\delta = 0.3$.}
   \label{fig:tauws}
\end{figure}Figure~\ref{fig:tauws} shows the temporal behaviour of the streamwise component of the shear stress at $\theta = \pi/2$ (outer wall) for $\Rey = 500$, $\delta = [0.01, 0.1, 0.3]$, $Q = 0.1$ and different values of the Womersley number. The values are normalised with the streamwise shear stress in the steady case at the same azimuthal position. Note that the azimuthal component of $\tauvec_w$ is zero at the outer and inner walls since the flow is symmetric with respect to the equatorial plane of the torus. Because the phase lag between the forcing and the flow field depends on the Womersley number, the signals of $\tau_{w s}$ at different values of $\Wo$ are not in phase with each other. Moreover, the amplitude of the streamwise shear stress does not exhibit a monotonic behaviour with the pulsation frequency. Indeed, for $1 \lessapprox \Wo \lessapprox 10$, the fluctuations of $\tau_{w s}$ decrease with increasing Womersley number, whereas they increase for $\Wo \gtrapprox 20$. The time-averaged value (dashed line in Figure~\ref{fig:tauws}) is approximately equal to the steady one for all the considered curvatures and Womersley numbers. However, for high frequencies, the amplitude of the fluctuations can reach almost $80\%$ of the steady value. This is particularly relevant from a structural point of view since it implies that the pipe is subject to highly unsteady loads. The effect of the relative amplitude of the unsteady component of the flow rate on the shear stress at the outer wall is reported in Figure~\ref{fig:tauws_Wo40} for $\Rey = 500$, $\delta = [0.01, 0.1, 0.3]$ and $\Wo = 40$. Since the phase lag does not exhibit significant variations with $Q$, the shown time signals are all in phase with each other. Increasing the pulsation amplitude, larger fluctuations of the streamwise wall shear stress are observed, following a monotonic trend. The amplitude of the fluctuations decreases monotonically with the curvature for a given pulsation amplitude. Indeed, for the toroidal pipe with $\delta = 0.01$, intracyclic variations of almost $250\%$ of the steady value are observed, whereas lower fluctuations occur for the other two investigated curvatures. For all investigated curvatures, negative values of the streamwise wall shear stress are reached within the cycle for a certain range of pulsation amplitudes, highlighting the occurrence of flow reversal. The flow at $\delta = 0.01$ is found to be the most prone to flow reversal at the investigated Reynolds number since backflow occurs for pulsation amplitudes greater than $Q = 0.2$. Note that, in all the considered cases, the time-averaged value (black dashed line in the figures) is very close to the steady one. \newline
The effect of the curvature on the wall shear stress changes as $\Wo$ varies, as it can be noticed in Figure~\ref{fig:tauws_Wo5_Q0p4}, which shows the intracyclic oscillations of $\tau_{w s}$ at the outer wall for $\Wo = 5$, $Q = 0.4$ and three different curvatures. Opposed to the case at $\Wo = 40$, a slight time shift is observed between the signals, since the phase lag displays a large variation with the curvature for this value of the Womersley number. Moreover, the amplitude of the fluctuations does not exhibit a monotonic trend with the curvature, and no symmetry with respect to the steady value is observed. At this value of the Womersley number, no flow reversal occurs for any of the considered curvatures, and the fluctuations around the steady value are considerably reduced with respect to the case at the same pulsation amplitude and $\Wo = 40$. 
\newline
For a better understanding of the unsteady loads acting on the toroidal pipe, the force balance can be investigated, as done by \citet{uchida1956} for the pulsatile flow in a straight pipe. Newton's second law in the streamwise direction reads
\begin{equation}
    \begin{split}
        \int_{s_1}^{s_2} \int_{0}^{2 \, \pi} \int_{0}^{R_p} \dfrac{F}{h_s} \, r \:\! h_s \:\! \dif r \, \dif \theta \, \dif s = & \int_{s_1}^{s_2} \int_{0}^{2 \, \pi} \tau_{w s} \:\! R_p \:\! h_s \:\! \dif \theta \, \dif s \ + \\
        & \int_{s_1}^{s_2} \int_{0}^{2 \, \pi} \int_{0}^{R_p} \dfrac{\partial U_s}{\partial t} \, r \:\! h_s \:\! \dif r \, \dif \theta \, \dif s \ + \\
        & \int_{s_1}^{s_2} \int_{0}^{2 \, \pi} \int_{0}^{R_p} \mathcal{C}(\Uvec) \, r \:\! h_s \:\! \dif r \, \dif \theta \, \dif s,
    \end{split}
\end{equation}
where $\mathcal{C}(\Uvec)$ is a shorthand for the (non-linear) advective terms in the Navier--Stokes equations. The force balance for $\Rey = 500$, $\Wo = 10 $, $Q = 0.3$, $\delta = 0.3$ is shown in Figure~\ref{fig:force_bal_d0p3_Wo10_Q0p3}. A black dashed line indicates the forcing amplitude for the laminar steady flow at the same Reynolds number and curvature. Note that a measure of the power that needs to be supplied to the system is given by the absolute value of the forcing amplitude $F$. Since the friction force is always directed against the main motion, it plays a beneficial role in reducing the needed power for a negative streamwise acceleration ($0 \lessapprox t/T \lessapprox 0.5$). Indeed, in this portion of the cycle, the required streamwise forcing is lower than the one in the steady case. An opposite situation is observed when the acceleration is positive ($0.5 \lessapprox t/T \lessapprox 1$) since the volume force has to counteract the friction and accelerate the flow, resulting in a larger value than the steady one. The flow at $\Wo = 10$ represents an intermediate case. Indeed, for lower frequencies, the acceleration term is small and the required forcing is mainly counteracting the friction, whereas for higher values of the Womersley number, the volume force primarily provides an acceleration of the flow in the streamwise direction. Note that the mean value of the streamwise forcing is equal to $F^{(0)}$, and its increase over the steady value varies quadratically with the pulsation amplitude, as previously discussed.

\subsection{Flow regimes}\label{sec:flow_regimes}
\begin{figure}
   \hspace{-0.45cm}
   \subfloat{\includegraphics[]{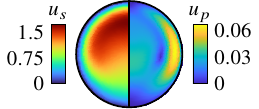}}   
   \subfloat{\includegraphics[]{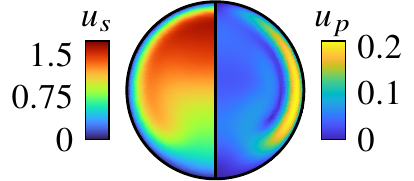}}   
   \subfloat{\includegraphics[]{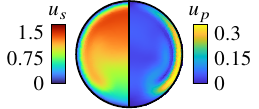}} 
   \caption{Laminar steady flow at $\Rey = 500$ and (\textit{left}) $\delta = 0.01$, (\textit{middle}) $\delta = 0.1$, (\textit{right}) $\delta = 0.3$. The streamwise velocity component is shown on the left side of the cross-section, whereas the right side displays the in-plane velocity magnitude. The inner wall of the bend is located on the bottom, whereas the top corresponds to the outer wall.}
   \label{fig:steady_base}
\end{figure}

\begin{figure}
   \hspace{-0.2cm}
   \subfloat{\includegraphics[]{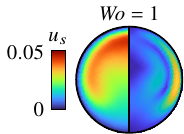}} \hskip 2pt  
   \subfloat{\includegraphics[]{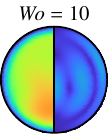}} \hskip 2pt  
   \subfloat{\includegraphics[]{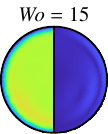}} \hskip 2pt  
   \subfloat{\includegraphics[]{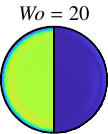}} \hskip 2pt  
   \subfloat{\includegraphics[]{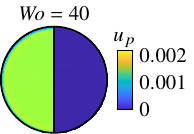}}  \\ \vskip 2pt 
   \hspace{-0.2cm}
   \subfloat{\includegraphics[]{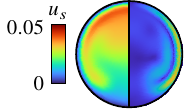}} \hskip 2pt
   \subfloat{\includegraphics[]{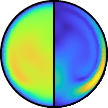}} \hskip 2pt  
   \subfloat{\includegraphics[]{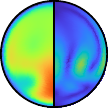}} \hskip 2pt 
   \subfloat{\includegraphics[]{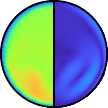}} \hskip 2pt   
   \subfloat{\includegraphics[]{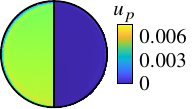}} \\ \vskip 2pt 
   \hspace{-0.2cm}
   \subfloat{\includegraphics[]{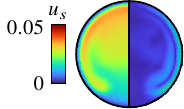}} \hskip 2pt
   \subfloat{\includegraphics[]{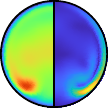}} \hskip 2pt  
   \subfloat{\includegraphics[]{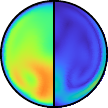}} \hskip 2pt
   \subfloat{\includegraphics[]{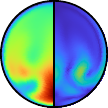}} \hskip 2pt  
   \subfloat{\includegraphics[]{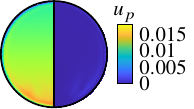}} 
   \caption{Real part of the fundamental frequency component ($n = 1$) of the laminar pulsatile flow at $\Rey = 500$, $Q = 0.05$,  $\Wo = [1, 10, 15, 20, 40]$. (\textit{Top}) $\delta = 0.01$, (\textit{middle}) $\delta = 0.1$, (\textit{bottom}) $\delta = 0.3$. The left side of the cross-section displays the streamwise velocity component, whereas the in-plane velocity magnitude is shown on the right side. The outer wall of the pipe corresponds to the top, whereas the inner one is located at the bottom.}
   \label{fig:vel_unsteady}
\end{figure}

The variation of the topology of the steady solution with the curvature has been previously investigated by \citet{canton2017}. The streamwise velocity component $u_s$ and the in-plane velocity magnitude $u_{i \:\!\! p} = (u_r^2 + u_{\theta}^2)^{1/2}$ of the laminar steady flows at $\Rey = 500$ and $\delta = [0.01, 0.1, 0.3]$ are presented in Figure~\ref{fig:steady_base} for reference. The steady component is virtually unaffected by the pulsation in the considered parameter range. The real part of the fundamental unsteady component ($n = 1$) is instead displayed in Figure~\ref{fig:vel_unsteady} for $\Rey = 500$, $Q = 0.05$, $\Wo = [1, 10, 15, 20, 40]$ and the three considered curvatures. It represents the contribution to the total flow field at $t = 0$ and $t = T/2$.\newline
The following is a qualitative description of the flow regimes throughout the parameter space. In the considered parameter range, it is observed that the structure of the flow fields does not change noticeably between the solutions at $K_b = 4$ and $K_b = 6$. Therefore, only flow fields at $K_b = 4$ are shown for the qualitative discussion, allowing the study of a broader parameter space at a reasonable computational cost.
\newline
A comparison between Figure~\ref{fig:steady_base} and Figure~\ref{fig:vel_unsteady} provides useful information about the variation of the flow topology at different Womersley numbers. For all curvatures, a quasi-steady regime can be observed for $\Wo \lessapprox 1$. Indeed, the real part of the unsteady component is similar to the steady one. Hence, the total flow field has a topology very close to that of the steady case, but with a time-periodic flow rate. At $\Wo \approx 10$, the flow starts to deviate from the quasi-steady behaviour for all investigated curvatures, marking the beginning of the intermediate regime. The flow topology becomes more complex for these values of the Womersley number because the pulsation time scale is similar to the intrinsic flow time scale. The streamwise velocity component exhibits larger values in the region between the inner wall and the core of the pipe for $\delta = 0.01$, whereas an increase of $u_s$ close to the walls on the lateral side of the cross-section is observed for the other two curvatures. However, the in-plane velocity magnitude does not display a large variation of its spatial structure with respect to the steady flow. For $\delta = 0.1$ and $\delta = 0.3$, the intermediate regime becomes more prominent for values of the Womersley number between $15$ and $20$, with the streamwise velocity increasing in the core of the pipe, towards the inner bend. The in-plane velocity magnitude of the first unsteady component has a peak close to the centre of the cross-section, indicating a considerable modification of the cross-flow topology because of the pulsation. However, for the lowest curvature, a different behaviour is observed, referred to as plug-flow regime, starting at $Wo \approx 15$ upwards. The streamwise velocity of the first unsteady component is almost uniform throughout the cross-section and a thin shear layer forms close to the pipe wall, whereas very low values of the in-plane velocity magnitude are observed. Away from the wall, the total flow field has the same spatial structure as the corresponding steady case but shifted in amplitude. As the Womersley number increases, the plug-flow behaviour becomes more marked. For the other two curvatures, the occurrence of the plug-flow regime starts at higher Womersley numbers than for $\delta = 0.01$. Indeed, for $\delta = 0.1$, the flow at $\Wo = 20$ lies between the intermediate and the plug-flow regimes, the latter becoming more clear for $\Wo \gtrapprox 25$. For $\delta = 0.3$, the intermediate regime extends up to $\Wo \approx 35$ and the flow exhibits a purely plug-flow behaviour for values of the Womersley number greater than $50$.

\section{Conclusions and outlook} \label{sec:concl}
A novel methodology for computing laminar solutions for the pulsatile flow in a toroidal pipe is presented in the present work. The Navier--Stokes equations are expressed in the frequency domain so that the laminar time-periodic solution can be computed as a fixed point of the system. A Newton--Raphson algorithm is used for this purpose. The method represents an alternative to the standard time-stepping approach. With the formulation in frequency space, also unstable periodic orbits can be computed, for which a stabilisation technique would be required in a time-stepping method. Moreover, any explicit dependence on both the intrinsic time scale of the flow and the pulsation period is removed in the fixed-point iteration approach. Therefore, the present fixed-point iteration and the time-stepper method complement each other, increasing the accessible parameter range for the pulsatile flow through toroidal pipes. The developed approach exhibits the usual limitations of algorithms based on the Newton--Raphson method, for which only local convergence is guaranteed. Moreover, in case of strong non-linear interactions, a large number of Fourier components need to be retained in the solution of the system, increasing the computational cost of the algorithm.\newline
The fixed-point iteration algorithm is used to compute laminar pulsatile solutions for $\Rey = 500$, $\delta = [0.01, 0.1, 0.3]$ and several values of the pulsation amplitude $Q$ and the Womersley number $\Wo$. It is pointed out that a purely sinusoidal input does result in an output with a broadband frequency content. Therefore, the prescription of either the flow rate or the pressure gradient plays a primary role in the definition of the problem. In the current investigation, a purely sinusoidal volume force, mimicking the effect of a streamwise pressure gradient, is imposed. A phase lag is observed between the mass flow rate and the volume forcing. It is found to be dependent on both the curvature and the Womersley number, whereas very small variations with the pulsation amplitude are reported. It reaches the asymptotic value of 90\textdegree\ as $\Wo$ increases, as in the case of the pulsatile flow in a straight pipe \citep{uchida1956}. Moreover, for a given Womersley number, the phase lag increases monotonically as the curvature decreases, approaching the value for $\delta = 0$ at the same Womersley number. The steady component of the forcing increases quadratically with $Q$ with respect to the value in the steady case, although the dependence becomes weaker as the Womersley number increases. On the other hand, the unsteady component of the forcing increases linearly with the pulsation amplitude and the proportionality coefficient gets larger with $\Wo$. The analysis of the shear stress at the outer wall shows that, even though the time-averaged value is approximately equal to that in the steady case, large fluctuations can be experienced, the structure potentially undergoing largely unsteady loads. \newline
Eventually, a characterisation of the spatial structure of the fundamental unsteady component reveals the existence of three regimes, and the ranges of Womersley numbers at which they occur are identified. For low values of $\Wo$, the total flow field corresponds to a periodic modulation of the steady flow component, giving rise to the quasi-steady regime. On the other hand, at high frequencies, a plug-flow behaviour is observed, with the steady flow shifted in amplitude during the cycle. An intermediate regime is identified between these two, characterised by a complex spatial structure of the fundamental unsteady component. The extent of this regime in $\Wo$ is found to increase with the curvature. \newline
The present study introduces a new methodology for computing a laminar pulsatile solution in curved pipes, which can be employed to extract the base flow for performing Floquet stability analysis \citep{Kern2023}. Laminar pulsatile flows in toroidal pipes represent a simplification of more complex cases found in several industrial and biological applications at low Reynolds number. The study of this family of flows provides further understanding of the effect of the pulsation when a secondary flow occurs, paving the way for the investigation of pulsatile flows in spatially developing bends. Furthermore, the numerical code and the methodology presented in the current work can be extended to helical configurations for studying the effect of the torsion.
\section*{Acknowledgements}
Financial support by the Swedish Research Council (VR) through the grant no.~2017-04421 and by the European Research Council under grant agreement 694452-TRANSEP-ERC-2015-AdG is acknowledged.


\appendix
\section{Navier--Stokes equations in toroidal coordinates}\label{sec:nonlinear_equations}
This section presents the Navier–Stokes equations \eqref{eq:NSE} expressed in toroidal coordinates. Defining $h_s = 1 + \delta \:\! r \sin({\theta})$ and setting the streamwise pressure gradient to zero since the flow is driven by the volume force defined in equation~\eqref{eq:forcing_shape}, they can be written as 
\begin{subequations}
\begin{align}
\begin{split}
    &\svec\textit{\textbf{-momentum:}}\\[1pt]
	&\frac{\partial u_s}{\partial t} + \frac{1}{h_s}\:\! \frac{\partial ({u_s} \:\! {u_s})}{\partial s} + \frac{1}{h_s \:\! r}\:\! \frac{\partial (h_s \:\! r \:\! {u_s} \:\! {u_r})}{\partial r} + \frac{1}{h_s \:\! r} \:\! \frac{\partial (h_s \:\! {u_s} \:\! {u_{\theta}})}{\partial {{\theta}}} + \frac{\delta \sin({\theta})}{h_s} \:\! {u_s} \:\! {u_r} \\[5pt]
	&  + \frac{\delta \cos({\theta})}{h_s} \:\! {u_s} \:\! {u_{\theta}} - \frac{F}{h_s} - \frac{1}{\Rey}\left[\frac{2}{h_s}\frac{\partial} {\partial s}\left(\frac{1}{h_s}\left(\frac{\partial {u_s}}{\partial s} + \delta \sin({\theta}) \:\! {u_r} + \delta \cos({\theta}) \:\! {u_{\theta}}\right)\right) \right. \\[5pt] &\left.
	+ \frac{1}{h_s \:\! r} \:\! \frac{\partial}{\partial r}\left(h_s^2 \:\! r \frac{\partial}{\partial r}\left(\frac{{u_s}}{h_s}\right) + r \:\! \frac{\partial {u_r}}{\partial s}\right) + \frac{1}{h_s \:\! r} \:\! \frac{\partial}{\partial {{\theta}}}\left(\frac{h_s^2}{r} \:\! \frac{\partial}{\partial {{\theta}}}\left(\frac{{u_s}}{h_s}\right) + \frac{\partial {u_{\theta}}}{\partial _s}\right)
	\right. \\[5pt] &\left.
	+ \:\! \delta \sin({\theta}) \left(\frac{\partial}{\partial r}\left(\frac{{u_s}}{h_s}\right) + \frac{1}{h_s^2} \:\! \frac{\partial {u_r}}{\partial s}\right) + \delta \cos({\theta}) \left(\frac{1}{r} \:\! \frac{\partial}{\partial {{\theta}}}\left(\frac{{u_s}}{h_s}\right)+\frac{1}{h_s^2}\:\! \frac{\partial {u_{\theta}}}{\partial s}\right)\right] = 0;
\end{split}&\\[2pt]
\begin{split}
    &\rvec\textit{\textbf{-momentum:}}\\[1pt]
	&\frac{\partial u_r}{\partial t} + \frac{1}{h_s} \:\! \frac{\partial ({u_s} \:\! {u_r})}{\partial s} + \frac{1}{h_s \:\! r} \:\! \frac{\partial (h_s \:\! r \:\! {u_r} \:\! {u_r})} {\partial r}  + \frac{1}{h_s \:\! r} \:\! \frac{\partial (h_s \:\! {u_r} \:\! {u_{\theta}})} {\partial {{\theta} }} - \frac{\delta \sin({\theta}) }{h_s} \:\!{u_s} \:\! {u_s} - \frac{{u_{\theta}} \:\! {u_{\theta}}}{r} + \frac{\partial p} {\partial r} \\[5pt]
	& - \frac{1}{\Rey}\left[\frac{1}{h_s} \:\! \frac{\partial} {\partial s}\left(h_s \:\! \frac{\partial} {\partial r}\left(\frac{{u_s}}{h_s}\right) + \frac{1}{h_s} \:\! \frac{\partial {u_r}}{\partial s}\right) 
	+ \frac{2}{h_s \:\! r} \:\! \frac{\partial}{\partial r}\left(h_s \:\! r \:\! \frac{\partial {u_r}} {\partial r}\right) + \frac{1}{h_s} \:\! \frac{\partial}{\partial {{\theta} }}\left(h_s\left(\frac{1}{r^2} \:\! \frac{\partial {u_r}}{\partial {{\theta}}} + \frac{\partial}{\partial r} \bigg(\frac{{u_{\theta}}}{r}\bigg)\right)\right) 
	\right. \\[5pt] &\left.
	- \frac{2 \:\! \delta \sin({\theta})}{h_s^2}\left(\frac{\partial {u_s}}{\partial s} + \delta \sin({\theta}) \:\! {u_r} + \delta \cos({\theta}) \:\! {u_{\theta}}\right) - \frac{2}{r^2}\left(\frac{\partial {u_{\theta}}} {\partial {\theta}} + {u_r}\right)\right] = 0;
\end{split}&\\[2pt]
\begin{split}
    &{\thetavec}\textit{\textbf{-momentum:}}\\[1pt]
	&\frac{\partial u_{\theta}}{\partial t} + \frac{1}{h_s} \:\! \frac{\partial ({u_s} \:\! {u_{\theta}})}{\partial s} + \frac{1}{h_s \:\! r} \:\! \frac{\partial (h_s \:\! r \:\! {u_r} \:\! {u_{\theta}})} {\partial r} + \frac{1}{h_s \:\! r} \:\! \frac{\partial (h_s \:\! {u_{\theta}} \:\! {u_{\theta}})}{\partial {{\theta}}} -\frac{\delta \cos({\theta})}{h_s} \:\! {u_s} \:\! {u_s} + \frac{{u_r} \:\! {u_{\theta}}}{r} + \frac{1}{r} \:\! \frac{\partial p}{\partial {{\theta}}} \\[5pt]
	&- \frac{1}{\Rey}\left[- \frac{2 \:\! \delta \cos({\theta})}{h_s^2}\left(\frac{\partial {u_s}}{\partial s} + \delta \sin({\theta}) \:\! {u_r} + \delta \cos({\theta}) \:\! {u_{\theta}}\right)\right. 
	+ \frac{1}{h_s \:\! r} \:\! \frac{\partial} {\partial r}\left(h_s\left(\frac{\partial {u_r}}{\partial {{\theta}}} + r^2 \:\! \frac{\partial}{\partial r}\bigg(\frac{{u_{\theta}}}{r}\bigg)\right)\right) \\[5pt] &\left. 
    + \frac{2}{h_s \:\! r^2} \:\! \frac{\partial}{\partial{{\theta}}}\left(h_s\left(\frac{\partial {u_{\theta}}}{\partial {{\theta}}} + {u_r}\right)\right)
    + \frac{1}{h_s} \:\! \frac{\partial}{\partial s} \left(\frac{h_s}{r} \:\! \frac{\partial}{\partial {{\theta}}}\left(\frac{{u_s}}{h_s}\right) + \frac{1}{h_s} \:\! \frac{\partial {u_{\theta}}}{\partial s}\right)
	 + \left(\frac{1}{r^2} \:\! \frac{\partial {u_r}}{\partial {{\theta}}} + \frac{\partial}{\partial r}\bigg(\frac{{u_{\theta}}}{r}\bigg)\right) \right] = 0;
\end{split}&\\[2pt]
\begin{split}
    &\textit{\textbf{incompressibility constraint:}}\\[1pt]
    &\frac{\partial (r\:\! u_s)}{\partial s} + \frac{\partial (h_s \:\! r \:\! {u_r})}{\partial r} + \frac{\partial (h_s \:\! {u_{\theta}})}{\partial {{\theta}}} = 0.
\end{split}
\end{align}
\end{subequations}

\section{Operators for linearised Navier--Stokes equations in toroidal coordinates}\label{app:linear_operators}
This section presents the operators appearing in the Navier--Stokes equations linearised about a given state $\qvec = (\uvec, p)^T$ and expressed in toroidal coordinates. The operator $\mathcal{L}_1$ representing the linear terms can be written as 
\begin{equation}
\mathcal{L}_1 = 
\begin{pmatrix}
   \mathcal{L}_{1, s s} & \mathcal{L}_{1, s r} & \mathcal{L}_{1, s \theta} & \mathcal{L}_{1, s p} \\[2pt]
   \mathcal{L}_{1, r s} & \mathcal{L}_{1, r r} & \mathcal{L}_{1, r \theta} & \mathcal{L}_{1, r p}\\[2pt]
   \mathcal{L}_{1, \theta s} & \mathcal{L}_{1,\theta r} & \mathcal{L}_{1,\theta \theta} & \mathcal{L}_{1, \theta p} \\[2pt]
   \mathcal{L}_{1, p s} & \mathcal{L}_{1, p r} & \mathcal{L}_{1, p \theta} & 0 
\end{pmatrix},
\label{eq:lin1_op}
\end{equation}
where
\begin{subequations}
        \begin{flalign}
            \begin{split}
                \mathcal{L}_{1, s s} = & +\frac{1}{r^2 \:\! \Rey} \:\! \frac{\partial^2 }{\partial {\theta}^2} - \frac{\delta^2 }{\Rey \, (\delta \:\! r \sin({\theta}) + 1)^2}  +\frac{1}{\Rey} \:\! \frac{\partial^2 }{\partial r^2} 
                - \left(\frac{1}{r \:\! \Rey \left(\delta \:\! r \sin({\theta}) + 1\right)} -\frac{2}{r \:\! \Rey}\right) \frac{\partial}{\partial r} \\ &
                + \frac{\delta \cos({\theta})}{r \:\! \Rey \left(\delta \:\! r \sin({\theta}) + 1\right)} \:\! \frac{\partial}{\partial {\theta}} 
                +\frac{2}{\Rey \, (\delta \:\! r \sin({\theta}) + 1)^2} \:\!\frac{\partial^2 }{\partial s^2},
            \end{split}&\\[3pt]
\begin{split}
\mathcal{L}_{1, s r} = & - \frac{1}{r \:\! \Rey} \left(\frac{3}{(\delta \:\! r \sin({\theta}) + 1)^2} - \frac{4}{\delta \:\! r \sin({\theta}) + 1}\right) \frac{\partial}{\partial s} + \frac{1}{\Rey \left(\delta \:\! r \sin({\theta}) + 1\right)} \:\! \frac{\partial^2}{\partial s \:\! \partial r},
\end{split}\\[3pt]
\begin{split}
\mathcal{L}_{1, s \theta} = & +\frac{1}{r \:\! \Rey \left(\delta \:\! r \sin({\theta}) + 1\right)} \:\! \frac{\partial^2}{\partial s \:\! \partial {\theta}} +\frac{3 \:\! \delta \cos({\theta})}{\Rey \, (\delta \:\! r \sin({\theta}) + 1)^2} \:\! \frac{\partial}{\partial s},
\end{split}&\\[3pt]
\begin{split}
\mathcal{L}_{1, s p} = & - \frac{1}{\delta \:\! r \sin({\theta}) + 1}\:\! \frac{\partial }{\partial s},
\end{split}\\[5pt]
\begin{split}
\mathcal{L}_{1, r s} = & - \left(\frac{3}{r \:\! \Rey \left(\delta \:\! r \sin({\theta}) + 1\right)} - \frac{3}{r \:\! \Rey \, (\delta \:\! r \sin({\theta}) + 1)^2}\right)\frac{\partial }{\partial s} + \frac{1}{\Rey \left(\delta \:\! r \sin({\theta}) + 1\right)} \:\! \frac{\partial^2}{\partial s \:\! \partial r},
\end{split}\\[3pt]
\begin{split}
\mathcal{L}_{1, r r} = & - \frac{2}{r^2 \:\! \Rey \, (\delta \:\! r \sin({\theta}) + 1)^2} - \frac{4 \:\! \delta \sin({\theta}) }{r \:\! \Rey \left(\delta \:\! r \sin({\theta}) + 1\right)}  + \frac{1}{r^2 \:\! \Rey} \:\! \frac{\partial^2}{\partial {\theta}^2} - \left(\frac{2}{r \:\! \Rey \left(\delta \:\! r \sin({\theta}) + 1\right)} - \frac{4}{r \:\! \Rey}\right) \frac{\partial}{\partial r} \\ & + \frac{\delta \cos({\theta})}{r \:\! \Rey \left(\delta \:\! r \sin({\theta}) + 1\right)} \:\! \frac{\partial}{\partial {\theta}} + \frac{1}{\Rey \, (\delta \:\! r \sin({\theta})+1)^2} \:\! \frac{\partial^2}{\partial s^2} + \frac{2}{\Rey}\:\! \frac{\partial^2}{\partial r^2},
\end{split}&\\[3pt]
\begin{split}
\mathcal{L}_{1, r \theta} = & - \frac{3 \:\! \delta \:\! \cos({\theta})}{r \:\! \Rey \left(\delta \:\! r \sin({\theta}) + 1\right)} + \frac{2 \:\! \delta \cos({\theta})}{r \:\! \Rey \, (\delta \:\! r \sin({\theta}) + 1)^2} - \frac{3}{r^2 \:\! \Rey} \:\! \frac{\partial}{\partial {\theta}} + \frac{\delta \cos({\theta})}{\Rey \left(\delta \:\! r \sin({\theta}) + 1\right)} \:\! \frac{\partial}{\partial r} + \frac{1}{r \:\! \Rey} \:\! \frac{\partial^2}{\partial r \:\! \partial {\theta}},
\end{split}\\[3pt]
\begin{split}
\mathcal{L}_{1, r p} = & - \frac{\partial }{\partial r},
\end{split}\\[3pt]
\begin{split}
\mathcal{L}_{1, \theta s} = & - \frac{3 \:\! \delta \cos({\theta})}{\Rey \, (\delta \:\! r \sin({\theta})+1)^2} \:\! \frac{\partial}{\partial s} + \frac{1}{r \:\! \Rey \left(\delta \:\! r \sin({\theta}) + 1\right)} \:\! \frac{\partial^2}{\partial s \:\! \partial {\theta}},
\end{split}\\[3pt]
\begin{split}
\mathcal{L}_{1, \theta r} = & - \frac{1}{r^2 \:\! \Rey}\left(\frac{1}{\delta \:\! r \sin({\theta})+1} - 4\right) \frac{\partial}{\partial {\theta}} + \frac{2 \:\! \delta  \cos({\theta})}{r \:\! \Rey \, (\delta  \:\! r \sin({\theta})+1)^2} + \frac{1}{r \:\! \Rey}\frac{\partial^2}{\partial r \:\! \partial {\theta}},
\end{split}\\[3pt]
\begin{split}
\mathcal{L}_{1, \theta \theta} = & - \frac{2 \:\! \delta ^2}{\Rey \, (\delta \:\! r \sin({\theta}) + 1)^2} - \frac{1}{r^2 \:\! \Rey}\left(\frac{3}{\delta \:\! r \sin({\theta}) + 1} - \frac{2}{(\delta \:\! r \sin({\theta})+1)^2}\right) + \frac{2 }{r^2 \:\! \Rey} \:\! \frac{\partial^2}{\partial {\theta}^2} + \frac{1}{\Rey} \:\! \frac{\partial^2}{\partial r^2}
\\ & - \left(\frac{1}{r \:\! \Rey\left(\delta \:\! r \sin({\theta}) + 1\right)} - \frac{2}{r \:\! \Rey}\right) \frac{\partial}{\partial r} + \frac{2 \:\! \delta \cos({\theta})}{r \:\! \Rey\left(\delta \:\! r \sin({\theta}) + 1\right)} \:\! \frac{\partial}{\partial {\theta}} + \frac{1}{\Rey \, (\delta \:\! r \sin({\theta}) + 1)^2} \:\! \frac{\partial^2}{\partial s^2},
\end{split}&\\[3pt]
\begin{split}
\mathcal{L}_{1, \theta p} = & - \frac{1}{r} \:\! \frac{\partial}{\partial {\theta}},
\end{split}\\[3pt]
\begin{split}
\mathcal{L}_{1, p s} = & + r \:\! \frac{\partial}{\partial s},
\end{split}\\[3pt]
\begin{split}
\mathcal{L}_{1, p r} = & + (2 \:\! \delta  \:\! r \sin({\theta})+1) + r \left(\delta  \:\! r \sin({\theta}) + 1\right) \:\! \frac{\partial}{\partial r},
\end{split}\\[3pt]
\begin{split}
\mathcal{L}_{1, p \theta} = & + \delta  \:\! r \cos({\theta}) + (\delta  \:\! r \sin({\theta}) + 1) \:\! \frac{\partial }{\partial {\theta}}.
\end{split}
\end{flalign}
\end{subequations}
%
%
%
%
The operator $\mathcal{L}_2(\uvec)$, which expresses the linearised convective terms, yields
\begin{equation}
\mathcal{L}_2(\uvec) = 
\begin{pmatrix}
   \mathcal{L}_{2, s s}(\uvec) & \mathcal{L}_{2,s r}(\uvec) & \mathcal{L}_{2, s \theta}(\uvec) & \ 0 \\
   \mathcal{L}_{2, r s}(\uvec) & \mathcal{L}_{2,r r}(\uvec) & \mathcal{L}_{2, r \theta}(\uvec) & \ 0 \\
   \mathcal{L}_{2, \theta s}(\uvec) & \mathcal{L}_{2, \theta r}(\uvec) & \mathcal{L}_{2, \theta \theta}(\uvec) & \ 0 \\
   0 & 0 & 0 & \ 0 \\
\end{pmatrix},
\label{eq:lin2_op}
\end{equation}
where
\begin{subequations}
\begin{flalign}
\mathcal{L}_{2, s s}(\uvec) = & - \frac{2 \:\! \delta \cos({\theta}) \:\! {u_{\theta}}}{\delta \:\! r \sin({\theta}) + 1} - \frac{\partial u_r}{\partial r} - \frac{3}{r} \:\! u_r - \frac{1}{r}\:\! \frac{\partial u_{\theta}}{\partial {\theta}} + \frac{2 \:\! {u_r}}{r\left(\delta \:\! r \sin({\theta}) + 1\right)} - {u_r} \frac{\partial}{\partial r} - \frac{u_{\theta}}{r} \frac{\partial}{\partial {\theta}} - \frac{2 \:\! {u_s}}{\delta \:\! r \sin({\theta}) + 1} \:\! \frac{\partial}{\partial s}, 
\\[3pt]
%
\mathcal{L}_{2, s r}(\uvec) = & - \frac{\partial u_s}{\partial r} + \frac{2 \:\! {u_s}}{r \left(\delta \:\! r \sin({\theta}) + 1\right)} - \frac{3}{r} \:\! {u_s} - {u_s} \:\! \frac{\partial}{\partial r},
\\[3pt]
%
\mathcal{L}_{2, s \theta}(\uvec) = & - \frac{2 \:\! \delta \cos({\theta}) \:\! {u_s}}{\delta \:\! r \sin({\theta}) + 1} - \frac{1}{r} \:\! \frac{\partial   u_s}{\partial {\theta}} - \frac{{u_s}}{r} \:\! \frac{\partial}{\partial {\theta}}, 
\\[3pt]
%
\mathcal{L}_{2, r s}(\uvec) = & - \frac{{u_r}}{\delta \:\! r \sin({\theta}) + 1} \:\! \frac{\partial }{\partial s} - \frac{2 \:\! {u_s}}{r\left(\delta \:\! r \sin({\theta}) + 1\right)} + \frac{2}{r} \:\! {u_s},
\\[3pt]
%
\mathcal{L}_{2, r r}(\uvec) = & - 2 \:\! {u_r} \:\! \frac{\partial}{\partial r} - \frac{{u_s}}{\delta \:\! r \sin({\theta}) + 1} \:\! \frac{\partial}{\partial s} - \frac{u_{\theta}}{r} \:\! \frac{\partial}{\partial {\theta}} - 2 \:\! \frac{\partial u_r}{\partial r}  - \frac{1}{r} \:\! \frac{\partial u_{\theta}}{\partial {\theta}} 
- \frac{{u_r} \left(4 \:\! \delta \:\! r \sin({\theta}) + 2\right)}{r \left(\delta \:\! r \sin({\theta}) + 1\right)} - \frac{{u_{\theta}} \:\! \delta \cos({\theta})}{\delta \:\! r \sin({\theta}) + 1},
\\[3pt]
%
\mathcal{L}_{2, r \theta}(\uvec) = & - \frac{1}{r} \frac{\partial u_r}{\partial {\theta}}
- \frac{{u_r} \:\! \delta \:\! \cos({\theta})}{\delta \:\! r \sin({\theta}) + 1} + \frac{2}{r} \:\! {u_{\theta}} - \frac{u_r}{r} \:\! \frac{\partial}{\partial {\theta}},
\\[3pt]
%
\mathcal{L}_{2, \theta s}(\uvec) = & - \frac{{u_{\theta}}}{\delta \:\! r \sin({\theta}) + 1} \:\! \frac{\partial}{\partial s} + \frac{2 \:\! \delta \cos({\theta}) \:\! {u_s}}{\delta \:\! r \sin({\theta})+1},
\\[3pt]
%
\mathcal{L}_{2, \theta r}(\uvec) = & - \frac{\partial u_{\theta}}{\partial r} + \frac{{u_{\theta}}}{r\left(\delta \:\! r \sin({\theta}) + 1\right)} - \frac{3}{r} \:\! {u_{\theta}} - {u_{\theta}} \:\! \frac{\partial}{\partial r},
\\[3pt]
%
\mathcal{L}_{2, \theta \theta}(\uvec) = & -\frac{\partial u_r}{\partial r} - \frac{3}{r} \:\! {u_r} - \frac{2}{r} \:\! \frac{\partial u_{\theta}}{\partial {\theta}} + \frac{{u_r}}{r \left(\delta \:\! r \sin({\theta}) + 1\right)} - {u_r}\frac{\partial}{\partial r} - \frac{2}{r} \:\! {u_{\theta}} \:\! \frac{\partial}{\partial {\theta}}
- \frac{{u_s}}{\delta \:\! r \sin({\theta}) + 1} \:\! \frac{\partial}{\partial s} - \frac{2 \:\! \delta \cos({\theta}) \:\! {u_{\theta}}}{\delta \:\! r \sin({\theta}) + 1}.  
%
\end{flalign}
\end{subequations}

\section{Symmetrisation operators}
\label{sec:TTs}
For odd values of $n_{\theta s}$, the symmetrisation operators $\Tmatr \in \mathbb{R}^{2N \times N}$ and $\Tmatr_s \in \mathbb{R}^{N \times 2N}$ are defined as
\begin{equation}
     \Tmatr = \begin{bmatrix}
             \Tmatr_A(s) & & & \\
             & \Tmatr_A(s) & & \\
             & & \Tmatr_A(-s) & & \\
             & & & \Tmatr_A(s) \!\!\!
         \end{bmatrix}, \quad \Tmatr_s = \onematr_{n_r} \otimes \big[ \onematr_4 \otimes \Bmatr \big]
\end{equation}
with
\begin{equation}
     \Tmatr_k(s) = \begin{bmatrix}
             \onematr_{n_r} \otimes \Amatr(s) & & & \\
             & \onematr_{n_r} \otimes \Amatr(s) & & \\
             & & \onematr_{n_r} \otimes \Amatr(-s) & & \\
             & & & \onematr_{n_r} \otimes \Amatr(s) \end{bmatrix}
\end{equation}
and
\begin{equation}
     \Amatr(s) = \begin{bmatrix}
             \onematr_n         & \\
             s\cdot \Jmatr_n  & \\
                         & \onematr_m           \\
                         & s\cdot \Jmatr_m
         \end{bmatrix} \in \mathbb {R}^{n_{\theta} \times n_{\theta s}},
     \quad 
     \Bmatr = \begin{bmatrix}
             \onematr_n & \zeromatr_n & \\
                 & & \onematr_m & \zeromatr_m \\
            \end{bmatrix} \in \mathbb {R}^{n_{\theta s} \times n_{\theta}}
\end{equation}
where $n = \left \lceil{n_{\theta s}/2}\right \rceil $, $m = n - 1$, $\Jmatr_n$ is the backward identity matrix of order $n$ and $s = \pm 1$ denotes the chosen symmetry, symmetric or antisymmetric, respectively. Empty parts of the matrices are zeros. In the case of even $n_{\theta s}$, similar but more convoluted expressions for $\Tmatr$ and $\Tmatr_s$ can be derived.

\bibliographystyle{apsrev4-2}
\bibliography{thesis}

\end{document}